\documentclass[aps,twocolumn,pra,superscriptaddress,showpacs,tightenlines]{revtex4-1}
\usepackage [latin1]{inputenc}
\usepackage{amsmath}
\usepackage{graphicx}
\usepackage{color}
\usepackage{amsfonts}
\usepackage{txfonts}
\usepackage[colorlinks,citecolor=blue]{hyperref}
\begin{document}
\title{Optomechanical dynamics in the $\mathcal{PT}$- and broken-$\mathcal{PT}$-symmetric regimes}
\author{Hai Xu}
\affiliation{College of Physics and Electronic Engineering, Institute of Solid State Physics, Sichuan Normal University, Chengdu 610101, People's Republic of China}

\author{Deng-Gao Lai}
\email{denggaolai@foxmail.com}
\affiliation{Theoretical Quantum Physics Laboratory, RIKEN Cluster for Pioneering Research, Wako-shi, Saitama 351-0198, Japan}

\author{Yi-Bing Qian}
\affiliation{College of Physics and Electronic Engineering, Institute of Solid State Physics, Sichuan Normal University, Chengdu 610101, People's Republic of China}

\author{Bang-Pin Hou}
\email{bphou@sicnu.edu.cn}
\affiliation{College of Physics and Electronic Engineering, Institute of Solid State Physics, Sichuan Normal University, Chengdu 610101, People's Republic of China}

\author{Adam Miranowicz}
\affiliation{Theoretical Quantum Physics Laboratory, RIKEN Cluster for Pioneering Research, Wako-shi, Saitama 351-0198, Japan}
\affiliation{Institute of Spintronics and Quantum Information, Faculty of Physics, Adam Mickiewicz University, 61-614 Pozna\'{n}, Poland}

\author{Franco Nori}
\affiliation{Theoretical Quantum Physics Laboratory, RIKEN Cluster for Pioneering Research, Wako-shi, Saitama 351-0198, Japan}
\affiliation{RIKEN Center for Quantum Computing (RQC), 2-1 Hirosawa, Wako-shi, Saitama 351-0198, Japan}
\affiliation{Physics Department, The University of Michigan, Ann Arbor, Michigan 48109-1040, USA}

\begin{abstract}
We theoretically study the dynamics of typical optomechanical systems, consisting of a passive optical mode and an active mechanical mode, in the $\mathcal{PT}$- and broken-$\mathcal{PT}$-symmetric regimes. By fully analytical treatments for the dynamics of the average displacement and particle numbers, we reveal the phase diagram under different conditions and the various regimes of both $\mathcal{PT}$-symmetry and stability of the system. We find that by appropriately tuning either mechanical gain or optomechanical coupling, both phase transitions of the $\mathcal{PT}$-symmetry and stability of the system can be flexibly controlled. As a result, the dynamical behaviors of the average displacement, photons, and phonons are radically changed in different regimes. Our study shows that $\mathcal{PT}$-symmetric optomechanical devices can serve as a powerful tool for the manipulation of mechanical motion, photons, and phonons.
\end{abstract}
\maketitle

\section{Introduction}

Cavity optomechanics, which explores the radiation-pressure interaction between electromagnetic and mechanical systems, has attracted considerable attention both theoretically and experimentally in the past decades~\cite{KV,AK,KR}. Due to optomechanical interaction, many interesting phenomena have been shown, such as cooling of mechanical oscillators to their quantum ground states~\cite{cooling1,cooling2,cooling3,BM,Wilson-Rae2007PRL,Marquardt2007PRL,Genes2008PRA,LaiH,Sommer2019PRL,Sommer2020PRR,Lai2020PRARC,Lai2021PRA}, photon blockade~\cite{Rabl2011PRL,Nunnenkamp2011,Liao2012PRA,Liao2013PRA,Wang2015PRA1,Huang2018PRL,Li2019PR,Zou2019PRA,Liao2020PRA}, generation and transfer of squeezed light~\cite{SG,PY,LXY2014PRL,LiuH,Qin2018PRL}, measurements with a high precision within the standard quantum limit~\cite{LC,GP,TF}, optomechanically induced effects of: nonreciprocity~\cite{FS,MR,MP,WL}, transparency (OMIT)~\cite{WR,SA,HW,Agarwal2010PRA,Lai2020PRA1}, absorption (OMIA)~\cite{MC}, and amplification~\cite{ZH,MH}.

It is usually assumed in quantum mechanics that the Hamiltonian must be Hermitian in order to ensure that their eigenvalues are real and that the time evolution operator is unitary. However, for parity-time ($\mathcal{PT}$)-symmetry quantum mechanics~\cite{BB,BBJ,Bender}, the effective Hamiltonian of a quantum system can be non-Hermitian, which is useful to describe a quantum system interacting with its environment. Note that this generalized approach to quantum mechanics does
not lead to any violations of no-go theorems in standard quantum mechanics, including quantum information~\cite{Yu2019}. A phase transition from the $\mathcal{PT}$-symmetric regime to the broken-$\mathcal{PT}$-symmetric regime can occur, when the $\mathcal{PT}$-symmetric condition is broken, and some eigenvalues become complex~\cite{BBP,LW}. The phase transition between the two regimes has been observed experimentally using various gain-loss-balanced systems, such as $\mathcal{PT}$-symmetric waveguides~\cite{KG,GS}, active LRC circuits~\cite{SL}, and $\mathcal{PT}$-symmetric whispering-gallery microcavities~\cite{PO}.

As an emerging frontier, optical-$\mathcal{PT}$-symmetric optomechanical systems~\cite{Jing2014PRL,LXY2015PRL,EP3,PTOMS1,PTOMS2,PTOMS3,PTOMS4,PTOMS5,PTOMS6}, which are realized by coupling an active (gain) cavity to a passive (lossy) optomechanical cavity, have led to various unconventional phenomena, such as phonon lasers~\cite{Jing2014PRL,Jing2017PRAP,Zhang2018NP}, $\mathcal{PT}$-enhanced OMIT~\cite{Jing2015SR,Jing2017SR,Li2016SR}, $\mathcal{PT}$-induced amplification~\cite{He2019PRA}, and coherent perfect absorption~\cite{Zhang2017PRA,Feng2011Science,Zhong2019OL}. Compared to these steady-state behaviors of $\mathcal{PT}$-symmetric systems, their dynamics can provide a more versatile description of these systems.

So far, the dynamics of photons have been predicted in optical-$\mathcal{PT}$-symmetric systems consisting of two waveguides~\cite{AQ} or two coupled cavities~\cite{NT,SE}. Subsequently, the dynamical behavior of the mechanical resonators has been studied in mechanical-$\mathcal{PT}$-symmetric four-mode hybrid optomechanical systems~\cite{XL}. Despite these advances, the dynamics of a typical optomechanical system, consisting of a passive optical mode coupled to an active mechanical mode, in the $\mathcal{PT}$- and broken-$\mathcal{PT}$-symmetric regimes, and the phase diagram under different conditions, have not yet been revealed.

In this paper, we focus on a \emph{comparative study} of the dynamics of a typical optomechanical system, which consists of a passive optical mode and an active mechanical mode implemented by a mechanical gain, in the $\mathcal{PT}$- and broken-$\mathcal{PT}$-symmetric regimes. Note that the mechanical gain can be achieved by phonon lasing or by coupling the mechanical mode to another cavity mode driven with a blue-detuned driving field~\cite{LW,Zhang2018PRA}. In contrast to previous work~\cite{XL} investigating the dynamics of mechanical modes in four-mode hybrid optomechanical systems, the aim here is not only to study the dynamics of \emph{both optical and mechanical} modes by \emph{fully analytical} treatments in \emph{typical} optomechanical systems which have more fundamental properties, but also to reveal in detail the \emph{phase diagram} under different conditions.

We find that by appropriately adjusting either the effective optomechanical coupling or the mechanical gain, phase transitions can be clearly observed. We obtain the \emph{phase diagram} under different conditions and the \emph{various regimes} of both $\mathcal{PT}$-symmetry and stability of the system. Using our \emph{exact analytical solutions} of the average displacement and particle numbers, their dynamical behaviors in different regimes can be understood adequately. We find that the energy exchange between the cavity and the mechanical oscillator is rapid (slow) for the $\mathcal{PT}$ (broken-$\mathcal{PT}$)-symmetric regime. This opens up the prospect to manipulate the exchange velocity of the excitations using $\mathcal{PT}$-symmetric optomechanical systems.

Moreover, spontaneous generation of the number of particles is discussed not only when gain compensates loss, but also when gain is \emph{not} equal to loss. Finally, we also find that: (i) the average displacement and the average particle numbers \emph{approach their steady-state values} in the asymptotically stable regime, (ii) they increase \emph{exponentially} in the unstable regime, and (iii) the average displacement \emph{oscillates periodically} in the finite-time stable regime, but not asymptotically stable. Our study reveals that $\mathcal{PT}$-symmetric systems can be used for the control of mechanical motion, photons, and phonons.

The remainder of the paper is organized as follows: In Sec.~\ref{section:model} we obtain the master equation of the $\mathcal{PT}$-symmetric-like optomechanical system by using a linearization procedure, when the dissipation and gain rates of the system are phenomenologically considered, and the differential equations for the average values are obtained from the master equation. In Sec.~\ref{section:PT}, the $\mathcal{PT}$-symmetry and stability of the $\mathcal{PT}$-symmetric-like optomechanical system are investigated through a phase diagram. In Sec.~\ref{section:x}, the dynamics of the average displacement of the mechanical oscillator are investigated in different regimes for the $\mathcal{PT}$-symmetric-like optomechanical system. And the dynamics of the average particle numbers in different stability regimes for the system are considered in Sec.~\ref{section:n}. The effect of spontaneous generation of particles is also studied in this section. Conclusions are presented in Sec.~\ref{section:conclusion}.

\begin{figure}[tbp]
\centering
\includegraphics[width=0.45 \textwidth]{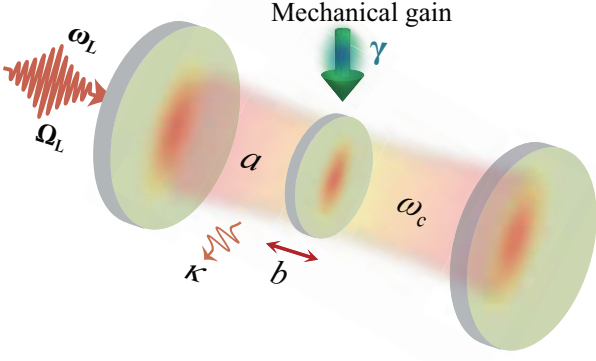}
\caption{Schematic of the $\mathcal{PT}$-symmetric-like optomechanical system, which consists of a passive cavity (with photon operator $a$, loss rate $\kappa$, and resonant frequency $\omega_{c}$), and an active mechanical oscillator (with phonon operator $b$, mechanical gain strength $\gamma$, and resonant frequency $\omega_{m}$). The cavity is driven by a control field with frequency $\omega_{L}$ and amplitude $\Omega_{L}$.}
\label{model0}
\end{figure}

\section{Model and equations of motion of average values}\label{section:model}

As schematically shown in Fig.~\ref{model0}, the considered optomechanical system consists of a passive cavity (with a loss rate $\kappa$) and an active mechanical oscillator (with a mechanical gain rate $\gamma$), which is called the $\mathcal{PT}$-symmetric-like optomechanical system~\cite{LW}. The cavity is driven by a control field with amplitude $\Omega_{L}=\sqrt{P_{L}\kappa/\hbar\omega_{L}}$, in which the input power and the frequency of the control field are given by $P_{L}$ and $\omega_{L}$, respectively.

The Hamiltonian of the system in the rotating reference frame at the frequency $\omega_{L}$ of the control field reads
\begin{align}
\hat{H}=&\;\hbar\Delta_{c} \hat{a}^{\dag}\hat{a}+\hbar\omega_{m}\hat{b}^{\dag}\hat{b}-\hbar g\hat{a}^{\dag}\hat{a}(\hat{b}+\hat{b}^{\dag})+i\hbar \Omega_{L}(\hat{a}^{\dag}-\hat{a}),\label{tH}
\end{align}
where $\hat{a}$ ($\hat{a}^{\dag}$) and $\hat{b}$ ($\hat{b}^{\dag}$) are the annihilation (creation) operators of the cavity field and the mechanical oscillator, respectively; $\omega_{m}$ is the resonance frequency of the mechanical oscillator, and $g$ is the single-photon optomechanical coupling strength. Moreover, $\Delta_{c}=\omega_{c}-\omega_{L}$ is the detuning between the cavity field of frequency $\omega_{c}$ and the control field of frequency $\omega_{L}$.

Due to the fact that the control field driving the cavity is strong, the Hamiltonian can be linearized by neglecting higher-order terms. Under the rotating-wave approximation (RWA), the linearized Hamiltonian is given by
\begin{align}
\hat{H}_{\mathrm{lin}}=\hbar\Delta\hat{a}^{\dag}\hat{a}+\hbar\omega_{m}\hat{b}^{\dag}\hat{b}-\hbar G(\hat{a}^{\dag}\hat{b}+\hat{a}\hat{b}^{\dag}),\label{lH}
\end{align}
where $\Delta=\Delta_{c}-g(\beta_{s}+\beta_{s}^{*})$ is the effective detuning between the cavity field and the control field, and $G=g\alpha_{s}$ is the effective optomechanical strength, and $\alpha_{s}$ and $\beta_{s}$ are the steady-state solutions of the system given by
\begin{align}
\alpha_{s}=\frac{\Omega_{L}}{i\Delta+\kappa}~ ~\mathrm{and}~ ~\beta_{s}=\frac{ig\alpha_{s}^{*}\alpha_{s}}{i\omega_{m}-\gamma}\label{s}.
\end{align}

The dynamics of the system can be described by the master equation in the Lindblad form, which is given by
\begin{align}
\frac{d}{dt}\rho=&\;\frac{1}{i\hbar}[H_{\rm lin},\rho]+
\kappa(2a\rho a^{\dag}-a^{\dag}a\rho-\rho a^{\dag}a)\notag\\
&+\gamma(2b^{\dag}\rho b-bb^{\dag}\rho-\rho bb^{\dag}).\label{ME}
\end{align}
The equations of motion of the mean values of an operator $\hat{o}$ can be calculated from the master equation in Eq.~(\ref{ME}) via $\frac{d}{dt}\langle\hat{o}\rangle=\mathrm{tr}(\hat{o}\dot{\hat{\rho}})$. Combining the commutation relations of operators $[\hat{i},\hat{j}^{\dag}]=\delta_{i,j}$, $[\hat{i},\hat{j}]=0$, and $[\hat{i}^{\dag},\hat{j}^{\dag}]=0$ ($\hat{i},\hat{j}=\hat{a},\hat{b}$), the equations of motion of $\langle\hat{a}\rangle$ and $\langle\hat{b}\rangle$ can be obtained as
\begin{subequations}
\begin{align}
&\frac{d}{dt}\langle\hat{a}\rangle=-i\Delta\langle\hat{a}\rangle
+iG\langle\hat{b}\rangle-\kappa\langle\hat{a}\rangle,\\
&\frac{d}{dt}\langle\hat{b}\rangle=-i\omega_{m}\langle\hat{b}\rangle
+iG\langle\hat{a}\rangle+\gamma\langle\hat{b}\rangle.
\label{eqad}
\end{align}
\end{subequations}
Correspondingly, the equations of motion of forms $\langle\hat{i}^{\dag}\hat{j}\rangle$ are given by
\begin{subequations}
\begin{align}
\frac{d}{dt}\langle \hat{a}^{\dag}\hat{b}\rangle&=(i\Delta-i\omega_{m}-\kappa+\gamma)\langle \hat{a}^{\dag}\hat{b}\rangle+iG(\langle \hat{a}^{\dag}\hat{a}\rangle-\langle \hat{b}^{\dag}\hat{b}\rangle),
\\
\frac{d}{dt}\langle \hat{a}^{\dag}\hat{a}\rangle&=-2\kappa\langle \hat{a}^{\dag}\hat{a}\rangle+iG\langle \hat{a}^{\dag}\hat{b}\rangle-iG\langle \hat{a}\hat{b}^{\dag}\rangle,
\\
\frac{d}{dt}\langle \hat{b}^{\dag}\hat{b}\rangle&=2\gamma\langle \hat{b}^{\dag}\hat{b}\rangle-iG\langle \hat{a}^{\dag}\hat{b}\rangle+iG\langle \hat{a}\hat{b}^{\dag}\rangle+2\gamma.\label{eqan}
\end{align}
\end{subequations}

\section{$\mathcal{PT}$-symmetry and stability}\label{section:PT}

\subsection{$\mathcal{PT}$-symmetry}

Taking the cavity loss rate $\kappa$ and the mechanical gain strength $\gamma$ into consideration, the effective Hamiltonian is obtained as
\begin{align}
\hat{H}_{\mathrm{eff}}=\;\hbar(\Delta-i\kappa)\hat{a}^{\dag}\hat{a}+\hbar(\omega_{m}+i\gamma)\hat{b}^{\dag}\hat{b}-\hbar G(\hat{a}^{\dag}\hat{b}+\hat{a}\hat{b}^{\dag}).\label{effH}
\end{align}
The properties of the space reflection (parity) operator $\mathcal{P}$ and the time reversal operator $\mathcal{T}$ are demonstrated as follows~\cite{BB,BBJ,Bender}. The action of the parity operator $\mathcal{P}$ on $\hat{H}_{\mathrm{eff}}$ is given by
\begin{align}
\mathcal{P}: \hat{a}\leftrightarrow -\hat{b},\ \hat{a}^{\dag}\leftrightarrow -\hat{b}^{\dag},
\end{align}
and the action of the time-reversal operator $\mathcal{T}$ on $\hat{H}_{\mathrm{eff}}$ is
\begin{align}
\mathcal{T}: \hat{a}\leftrightarrow \hat{a},\ \hat{a}^{\dag}\leftrightarrow \hat{a}^{\dag},\ \hat{b}\leftrightarrow \hat{b},\ \hat{b}^{\dag}\leftrightarrow \hat{b}^{\dag},\ i\leftrightarrow -i.\label{t}
\end{align}

After the combined actions of the parity and time-reversal operations, i.e., the $\mathcal{P}\mathcal{T}$ operations, the effective Hamiltonian in Eq.~(\ref{effH}) becomes
\begin{align}
\hat{H}_{\mathrm{eff}}^{\mathcal{PT}}=&\;\mathcal{PT}\hat{H}_{\mathrm{eff}}(\mathcal{PT})^{-1}\notag\\=
&\;\hbar(\Delta+i\kappa)\hat{b}^{\dag}\hat{b}
+\hbar(\omega_{m}-i\gamma)\hat{a}^{\dag}\hat{a}-\hbar G(\hat{a}^{\dag}\hat{b}+\hat{a}\hat{b}^{\dag}).\label{PTeffH}
\end{align}
From Eq.~(\ref{PTeffH}), we can obtain $H_{\mathrm{eff}}=H_{\mathrm{eff}}^{\mathcal{PT}}$ if and only if the relations $\Delta=\omega_{m}=\omega_{1}$ and $\kappa=\gamma$ are satisfied. In fact, the $\mathcal{PT}$-symmetry can be generalized to the case where the cavity decay rate $\kappa$ is not exactly equal to the mechanical gain strength $\gamma$. Therefore, in the latter case, only the relation $\Delta=\omega_{m}=\omega_{1}$ is always satisfied. The Hamiltonian of Eq.~(\ref{effH}) is rewritten as
\begin{align}
\hat{H}=\hbar\begin{pmatrix}\hat{a}^{\dag}&\hat{b}^{\dag}\end{pmatrix}
\begin{pmatrix}\omega_{1} -i\kappa &-G\\-G&\omega_{1}+i\gamma\end{pmatrix}\begin{pmatrix}\hat{a}\\\hat{b}\end{pmatrix}.
\label{mfH}
\end{align}
By diagonalizing the matrix in Eq.~(\ref{mfH}), the eigenfrequencies of the supermodes $\hat{A}_{\pm}=\left(\hat{a}\pm\hat{b}\right)$ can be obtained as
\begin{align}
\omega_{\pm}=\omega_{1}-\frac{i}{2}(\kappa-\gamma)\pm
\sqrt{G^{2}-\frac{1}{4}(\kappa+\gamma)^{2}}.
\end{align}

When $G>(\kappa+\gamma)/2$, the eigenfrequencies have two different real parts and an identical imaginary part, the system possesses the $\mathcal{PT}$-symmetry with two different frequencies and an identical linewidth, which is described by the regimes (2) and (4) in the phase diagram shown in Fig.~\ref{model}(a).

If the parameters satisfy the relation $G<(\kappa+\gamma)/2$, the eigenfrequencies have two different imaginary parts and an identical real part. The frequencies of the supermodes are the same, while their linewidths are different. Then the $\mathcal{PT}$-symmetry of the system is broken. The broken-$\mathcal{PT}$-symmetry corresponds to the regimes (1) and (3) in the phase diagram shown in Fig.~\ref{model}(a).

The phase transition of the $\mathcal{PT}$-symmetry takes place around the border point $G=(\kappa+\gamma)/2$, which is termed as an \emph{exceptional point} (EP)~\cite{EP4,EP5,EP6,AMandFN,AMandFN1,Lu2021PRA} as shown by the red line and blue point in the phase diagram. Note that this is a semiclassical EP, which corresponds to a
spectral degeneracy of a non-Hermitian Hamiltonian. The prediction
of a quantum EP would require the inclusion of quantum noise by
finding degeneracies of, e.g., a Liouvillian, as proposed
in Refs.~\cite{EP4,Minganti2020,AMandFN}.

\subsection{Stability}

The linearized equations of motion can be compactly written in a matrix form as
\begin{align}
\dot{\hat{u}}=A\hat{u},\label{de}
\end{align}
where $\hat{u}$ is the column vector of $\hat{u}^{T}=(\hat{a},\hat{a}^{\dag},\hat{b},\hat{b}^{\dag})$, and the square matrix $A$ is
\begin{align}
A=\begin{pmatrix}-i\omega_{1} -\kappa&0 &iG&0\\0&i\omega_{1} -\kappa&0 &-iG\\iG&0&-i\omega_{1}+\gamma&0\\0&-iG&0&i\omega_{1}+\gamma\end{pmatrix}.
\end{align}
The eigenvalues $\lambda$ of the matrix $A$ are
\begin{align}
\lambda_{{\rm \tau,s}}=\frac{1}{2}\left[\gamma-\kappa+{\rm \tau}\sqrt{(\gamma+\kappa)^{2}-4G^{2}}+i{\rm s}2\omega_{1}\right],\label{eoem}
\end{align}
where ${\rm \tau}=\pm 1$ and ${\rm s}=\pm 1$.

\begin{figure}[tbp]
\centering
\includegraphics[width=0.47 \textwidth]{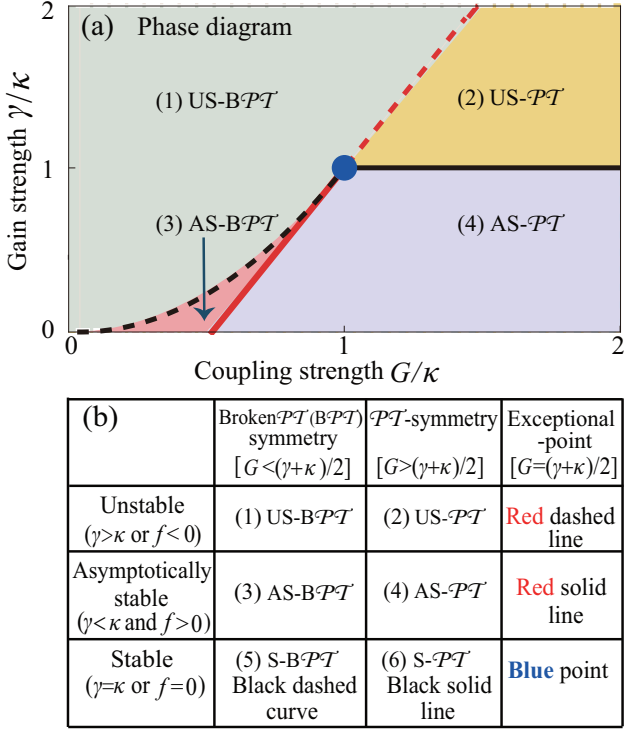}
\caption{(a) Phase diagram under different conditions of the mechanical gain rate $\gamma$ and the effective optomechanical strength $G=g\alpha_{s}$, in units of the cavity decay rate $\kappa$. There are two borders. The red line shows the border between the $\mathcal{PT}$-symmetric phase ($\mathcal{PT}$, on the right hand) and the broken-$\mathcal{PT}$-symmetric phase ($\mathrm{B}\mathcal{PT}$, on the left hand). The black dashed curve and black solid line are the border between the asymptotically stable (AS, below the border) phase and the unstable (US, above the border) phase. (b) Various regimes or regions of the $\mathcal{PT}$-symmetry and stability of the system. The new parameter $f$ in the table is defined as $f\equiv G^{2}-\gamma\kappa$.}
\label{model}
\end{figure}

The stability of the system can be discussed in the following cases~\cite{Braun,DK}:\\
(i) If the parameters satisfy the relations $f<0$ ($f=G^{2}-\gamma\kappa$) or $\gamma>\kappa$, some eigenvalues of $A$ have a positive real part, so the system is unstable. This corresponds to the regimes (1) and (2) in the phase diagram in Fig.~\ref{model}(a).
\\
(ii) When $f>0$ and $\gamma<\kappa$, all of the eigenvalues of $A$ have a negative real part, so the system lies in the asymptotically stable regime. This situation is described by the regimes (3) and (4) in Fig.~\ref{model}(a).
\\
(iii) When $f=0$ and $\gamma<\kappa$, the real parts of the two eigenvalues of $A$ are zero and those of the other two are negative; so the system is stable, and is described by the black dashed curve (5) in Fig.~\ref{model}(a).
\\
(iv) When $f>0$ and $\gamma=\kappa$, we find $\lambda_{{\rm \tau,s}}={\rm \tau}\sqrt{\kappa^{2}-G^{2}}+i{\rm s}\omega_1$. In this case, all the eigenvalues of $A$ have a vanishing real part and the corresponding four eigenvectors are linearly independent; thus, the system is in the finite-time stable regime, but not asymptotically stable, and shown by the black solid curve (6) in Fig.~\ref{model}(a).
\\
(v) When $f=0$ and $\gamma=\kappa$, the real parts of the eigenvalues of $A$ are zero and $A$ has only two linearly independent eigenvectors. In this case the system is unstable. This corresponds to the blue point in Fig.~\ref{model}(a).

\section{dynamis of the average displacement of the mechanical oscillator}\label{section:x}

\begin{figure*}[tbp]
\centering
\includegraphics[width=1 \textwidth]{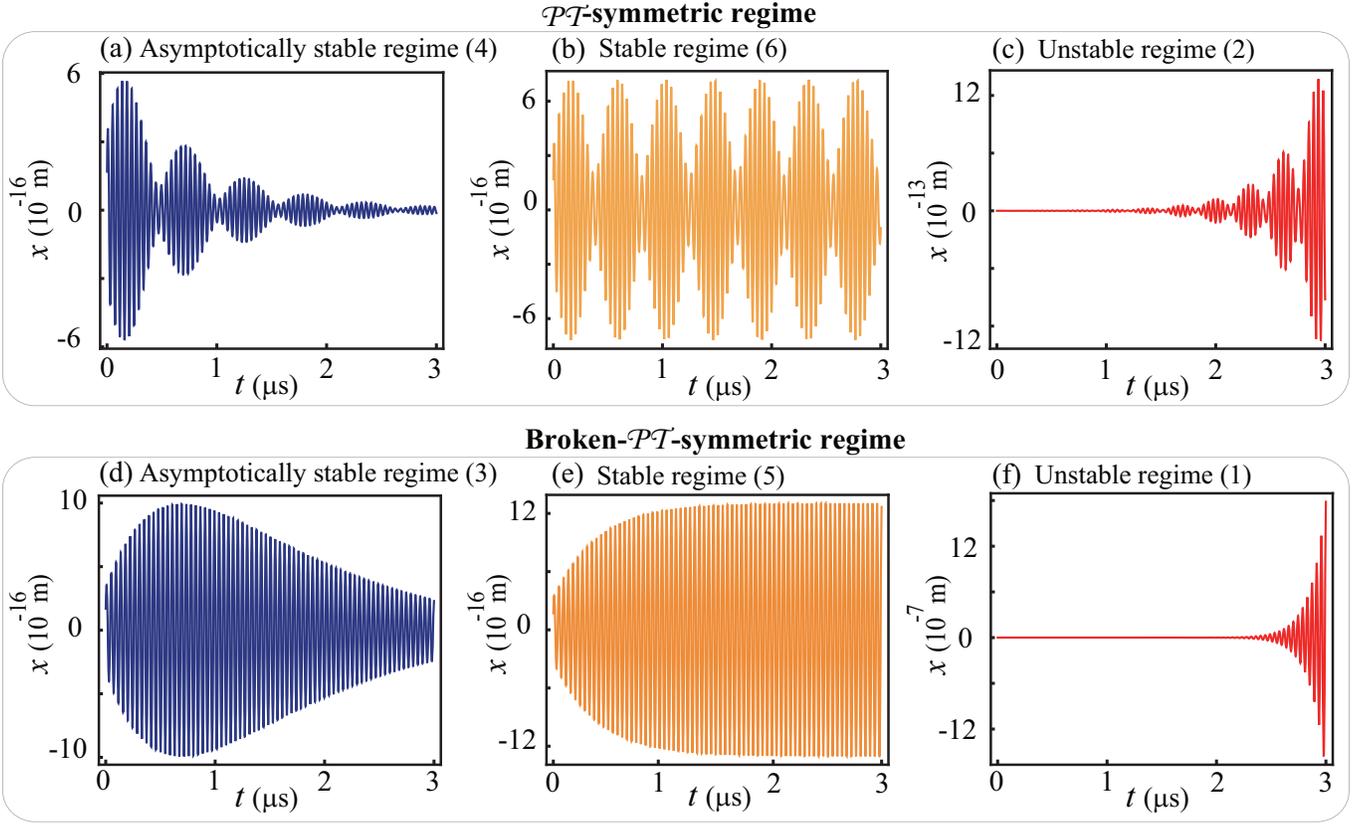}
\caption{Dynamics of the average displacement in the $\mathcal{PT}$-symmetric (first row) and broken-$\mathcal{PT}$-symmetric (second row) regimes. Examples for different stable regimes shown in Figs.~\ref{model}(b) and~\ref{model}(c): (a) and (d) correspond to the asymptotically stable regime, (b) and (e) correspond to the finite-time stable regime, but not asymptotically stable, (c) and (f) correspond to the unstable regime.
The gain rate $\gamma$ of the mechanical oscillator and the effective optomechanical coupling strength $G$ are given by: (a) $\gamma=0.6\kappa$ and $G=1.2\kappa$, (b) $\gamma=\kappa$ and $G=1.5\kappa$, (c) $\gamma=1.8\kappa$ and $G=2.1\kappa$, (d) $\gamma=0.6\kappa$ and $G=0.798\kappa$, (e) $\gamma=0.6\kappa$ and $G=\sqrt{0.6}\kappa$, (f) $\gamma=1.8\kappa$ and $G=1.2\kappa$.
Other parameters are set as: $\kappa=6.45{\rm MHz}$, $m=5\times10^{-11}{\rm kg}$, $\omega_{1}=23.4\times2\pi {\rm MHz}$~\cite{PO,EP3}, $\alpha=2\exp(i\pi/6)$, and $\beta=2\exp(i\pi/3)$.}
\label{ex}
\end{figure*}

Now, we consider the dynamics of the average displacement of the mechanical oscillator. Here, the initial state of the system is assumed to be a coherent state $|\alpha\rangle|\beta\rangle$, where the amplitudes of the coherent state are given, respectively, by $\alpha_{0}$ and $\beta_{0}$ with $\theta_{1}$ and $\theta_{2}$ being the initial phases.
The average value of the mechanical displacement, $x=\langle\hat{x}\rangle=\sqrt{\hbar/2m\omega_{1}}\big(\langle\hat{b}\rangle
+\langle\hat{b}\rangle^{*}\big)$, can be calculated by solving Eq.~(\ref{eqad}) in the case of $\Delta=\omega_{m}=\omega_{1}$ as
\begin{align}
x=&\;\frac{1}{\Omega}\sqrt{\frac{\hbar}{2m\omega_{1}}}\exp\left[\frac{1}{2}\left(\gamma-\kappa -2i\omega_{1}\right)t\right]\Big[\beta\Omega\cosh\left(
\frac{\Omega}{2}t\right)\notag\\
&+\left(\beta\gamma+\beta\kappa+2iG\alpha\right)\sinh\left(
\frac{\Omega}{2}t\right)\Big]+\rm{c.c.},\label{x}
\end{align}
where $\Omega=\sqrt{(\gamma+\kappa)^{2}-4G^{2}}$, which is an imaginary number in the $\mathcal{PT}$-symmetric regime [$G>(\kappa+\gamma)/2$], and the terms $\cosh\left(\frac{\Omega}{2}t\right)$ and $\sinh\left(\frac{\Omega}{2}t\right)$ are transformed into the form of a sinusoidal time function; while $\Omega$ is a real number in the broken-$\mathcal{PT}$-symmetric regime [$G<(\kappa+\gamma)/2$] and the expression can remain the same.

Based on the expression shown in Eq.~(\ref{x}), we investigate the dynamics of the mechanical displacement by plotting the time evolution of the average value of the displacement operator. First, we consider the dynamics of the average displacement in the $\mathcal{PT}$-symmetric regime. In Fig.~\ref{ex}(a), we set the parameters $\gamma=0.6\kappa$ and $G=1.2\kappa$ which lead the system to the asymptotically stable regime (4). We can see here that the oscillations of the displacement collapses and revivals with a decaying amplitude and asymptotically approach zero (at the equilibrium position) for a certain time.
When the values of the parameters are set as $\gamma=\kappa$ and $G=1.5\kappa$, as shown in Fig.~\ref{ex}(b), the system lies in the finite-time stable regime (6), but not asymptotically stable. It is shown here that the oscillations of the average displacement exhibit collapses and revivals periodically.
The dynamical behavior of the average displacement in the unstable regime (2) are displayed in Fig.~\ref{ex}(c) with the parameters given by $\gamma=1.8\kappa$ and $G=2.1\kappa$. It is shown that the average displacement of the mechanical oscillator oscillates with periodic collapses and revivals with increasing amplitude.

\begin{figure*}[tbp]
\centering
\includegraphics[width=1 \textwidth]{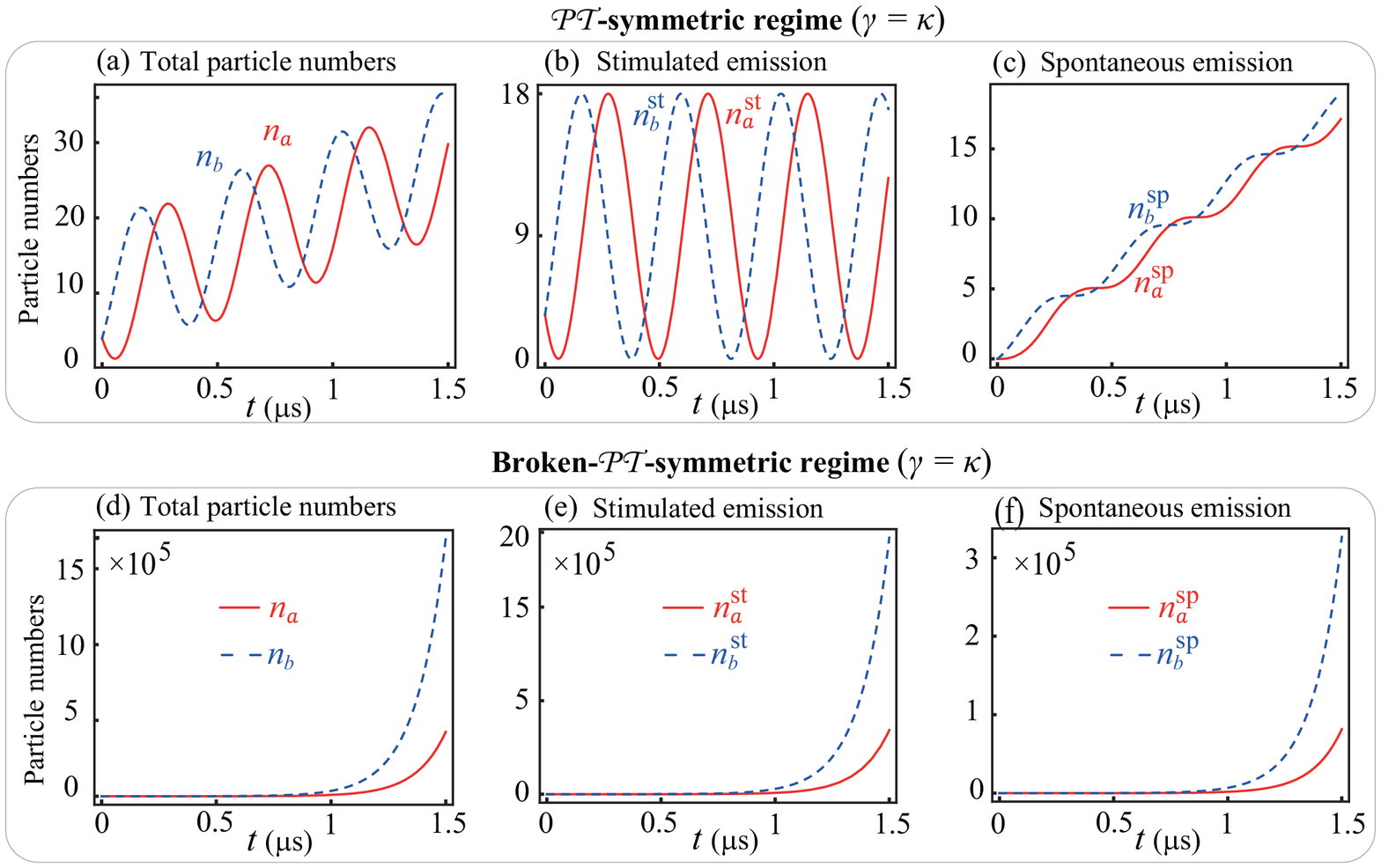}
\caption{The dynamics of the photon numbers (solid red curves) and phonon numbers (dashed blue curves) for the $\mathcal{PT}$-symmetric regime (first row with $G=1.5\kappa$) and the broken-$\mathcal{PT}$-symmetric regime (second row with $G=0.8$) in the case of $\gamma=\kappa$. The total average particle numbers $n_{a}$ and $n_{b}$ are given by (a) and (d); the numbers of particles generated by stimulated generation, $n_{a}^{\mathrm{st}}$ and $n_{b}^{\mathrm{st}}$, are given by (b) and (e); and those generated by spontaneous emission, $n_{a}^{\mathrm{sp}}$ and $n_{b}^{\mathrm{sp}}$, are given by (c) and (f), respectively. Other parameters are same as Fig.~\ref{ex}.}
\label{bn}
\end{figure*}

Second, we consider the dynamical evolution of the average displacement in the broken-$\mathcal{PT}$-symmetric regime. In Fig.~\ref{ex}(d), the parameters are set as $\gamma=0.6\kappa$ and $G=0.798\kappa$ which enables the system to be in the asymptotically stable regime (3). It is shown that the oscillations of the average displacement increases with time and then decreases to the equilibrium value $0$.
When the parameters are given by $\gamma=0.6\kappa$ and $G=\sqrt{0.6}\kappa$, as shown in Fig.~\ref{ex}(e), the system is in the finite-time stable regime (5), the oscillation amplitude of the average displacement increases with time and then approaches the constant value,
\begin{align}
A_{s}=&\;\frac{2}{\kappa-\gamma}\sqrt{\frac{\hbar}{2m\omega_{1}}}
\Big[\kappa^{2}|\beta|^2+\kappa\gamma|\alpha|^2-i\kappa\sqrt{\kappa\gamma}(\alpha^{*}\beta-\beta^{*}\alpha)\Big]^{\frac{1}{2}}.
\end{align}
In Fig.~\ref{ex}(f), we consider the dynamical evolution of the average displacement in the unstable regime (1) with parameter $\gamma=1.8\kappa$ and $G=1.2\kappa$. In this regime (f), the average displacement oscillates with an increasing amplitude with time.

By comparing the dynamics in the $\mathcal{PT}$-symmetric regime, shown in Figs.~\ref{ex}(a),~\ref{ex}(b), and~\ref{ex}(c) with those in the broken-$\mathcal{PT}$-symmetric regime, shown in Figs.~\ref{ex}(d),~\ref{ex}(e), and~\ref{ex}(f), we can see that the periodic collapses and revivals appear in the former case, while does not exist in the latter case.
In the three stable regimes, the different types of the dynamical behavior exhibit amplitude oscillations with time. Specifically, the oscillation amplitude of the average displacement $x$ decreases to $0$ as $t\rightarrow\infty$ when the system is asymptotically stable [regimes (3) and (4)]. However, the oscillation amplitude exponentially grows in the unstable regimes (1) and (2). While it periodically oscillates, with a constant amplitude, when the system is finite-time stable, but not asymptotically stable [regimes (5) and (6)]. These results open up an avenue to the manipulation of the mechanical motion by utilizing $\mathcal{PT}$-symmetric optomechanical devices.

\section{dynamics of the average particle numbers}\label{section:n}

\begin{figure*}[tbp]
\centering
\includegraphics[width=0.93 \textwidth]{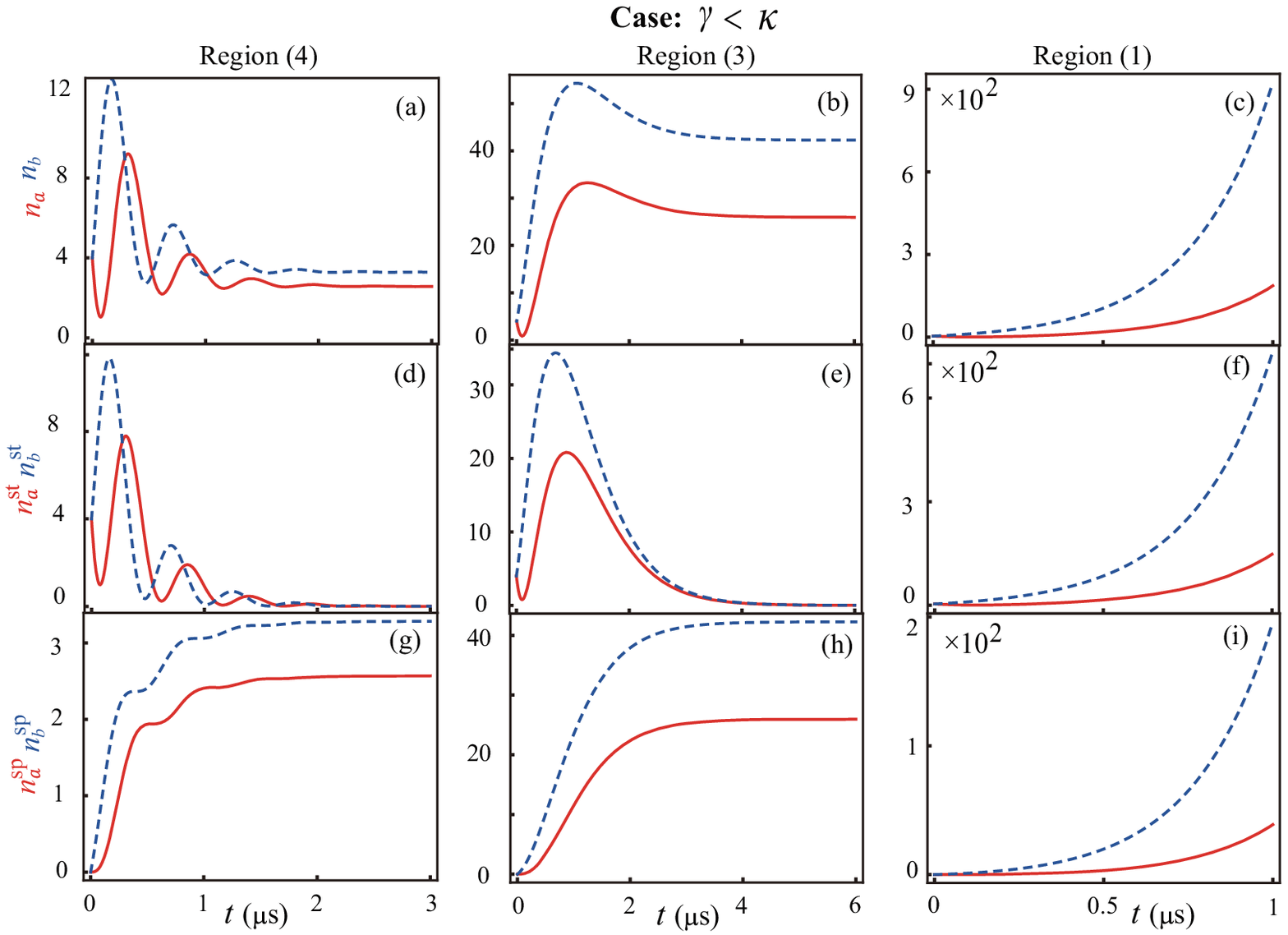}
\caption{Dynamics of photons (solid red curves) and phonons (dashed blue curves); (a-c) shows the total average particle numbers $n_{a}$ and $n_{b}$; (d-f) those generated by stimulated emission, $n_{a}^{st}$ and $n_{b}^{st}$; and (g-i) these photons and phonons generated by spontaneous emission, $n_{a}^{\mathrm{sp}}$ and $n_{b}^{\mathrm{sp}}$, when $\gamma\ < \kappa$. Here we assumed different values of the gain rate $\gamma$ and the effective optomechanical coupling strength $G$: (a,d,g) $\gamma=0.6\kappa$ and $G=1.2\kappa$, (b,e,h) $\gamma=0.6\kappa$ and $G=0.798\kappa$, (c,f,i) $\gamma=0.6\kappa$ and $G=0.6\kappa$. Other parameters are same as Fig.~\ref{ex}.}
\label{ln}
\end{figure*}

In the following, we discuss the dynamics of the average particle numbers in terms of the photon number $n_{a}=\langle\hat{a}^{\dag}\hat{a}\rangle$ and the phonon number $n_{b}=\langle\hat{b}^{\dag}\hat{b}\rangle$ by solving Eq.~(\ref{eqan}) evolving from a coherent state $|\alpha\rangle|\beta\rangle$ under the condition $\Delta=\omega_{m}=\omega_{1}$.
In order to understand the source of the generated particles more clearly, we divide the total average particle numbers $n_{i}$ ($i=a,b$) into two parts
\begin{align}
n_{i}=n_{i}^{\mathrm{st}}+n_{i}^{\mathrm{sp}},\label{n}
\end{align}
where $n_{i}^{\mathrm{st}}$ is the number of particles generated by stimulated emission, which depends on the initial values, and quantum noises are not considered. This part can be obtained from a semiclassical theory. The other term $n_{i}^{\mathrm{sp}}$ is the number of particles generated by spontaneous emission, which is induced by quantum noise~\cite{AQ,XL}.
We shall investigate the dynamics in the two cases $\gamma=\kappa$ and $\gamma\neq\kappa$, in which the expressions of the average numbers are different.

\subsection{Dynamics of the average particle numbers for $\gamma=\kappa$}

First, we consider the dynamics of the average numbers of particles, $n_{a}$ and $n_{b}$, in the case of $\gamma=\kappa$. The expressions of the photon numbers generated by stimulated emission, $n_{i}^{\mathrm{st}}$, and spontaneous emission, $n_{i}^{\mathrm{sp}}$, are given by

\begin{eqnarray}
n_{a}^{\mathrm{st}}&=&\frac{1}{4\Omega_{1}^{2}}\big[m_{1}+2o_{1}C_1+2o_{2}S_1)\big],
\notag\\
n_{b}^{\mathrm{st}}&=&\frac{1}{4\Omega_{1}^{2}}\big[m_{1}+2o_{3}C_1+2o_{4}S_1\big],
\notag\\
n_{a}^{\mathrm{sp}}&=&\frac{1}{4\Omega_{1}^{2}}\big[-4G^{2}\kappa t+2\frac{\kappa G^2}{\Omega_{1}}S_1\big],
\notag\\
n_{b}^{\mathrm{sp}}&=&\frac{1}{4\Omega_{1}^{2}}\left[-4G^{2}\kappa t+4\kappa^{2}\left(C_1-1\right)+2\left(\kappa\Omega_{1}+\frac{\kappa^{3}}{\Omega_{1}}\right)S_1\right],\label{spbnb}
\end{eqnarray}
where $C_1=\cosh\left(2\Omega_{1}t\right)$ and $S_1=\sinh\left(2\Omega_{1}t\right)$, with $\Omega_{1}=\sqrt{\kappa^{2}-G^{2}}$; $\Omega_{1}$ is imaginary in the $\mathcal{PT}$-symmetric regime, and the terms $C_{1}$ and $S_{1}$ transform into the form of a sinusoidal time function, while $\Omega_{1}$ is real in the broken-$\mathcal{PT}$-symmetric regime and the expression remains the same. Other coefficients are as follows,

\begin{eqnarray}
m_{1}&=&2i\kappa\delta-2G^{2}(|\alpha|^2+|\beta|^2),\notag\\
o_{1}&=&(\kappa^{2}+\Omega_{1}^{2})|\alpha|^2+G^{2}|\beta|^2-i\kappa\delta,\notag\\
o_{2}&=&i\Omega_{1}\delta-2\kappa\Omega_{1}|\alpha|^2,\notag\\
o_{3}&=&(\kappa^{2}+\Omega_{1}^{2})|\beta|^2+G^{2}|\alpha|^2-i\kappa\delta,\notag\\
o_{4}&=&-i\Omega_{1}\delta+2\kappa\Omega_{1}|\beta|^2,
\end{eqnarray}
where $\delta=G\left(\alpha^{*}\beta-\beta^{*}\alpha\right)$.

In the case of $\gamma=\kappa$, the system remains in the two regimes: (i) the finite-time stable regime, but not asymptotically stable and the $\mathcal{PT}$-symmetric regime, when the parameters satisfy the relation of $G>(\kappa+\gamma)/2$ [the regime (6)]; and (ii) the unstable and broken-$\mathcal{PT}$-symmetric regimes for $G<(\kappa+\gamma)/2$ [the regime (1)].

In the regime (6), it is shown in Figs.~\ref{bn}(a), ~\ref{bn}(b), and ~\ref{bn}(c) that the photon numbers (red solid curve) and phonons (blue dashed curve) oscillate periodically with a monotonously increasing equilibrium value. This is quite different from the dynamics around the constant value in the semiclassical theory (correspond to stimulated generation), and the phenomenon of the monotonically increasing photon and phonon numbers generated by spontaneous generation. The average particle numbers from spontaneous generation dominates the total generation of the average particle numbers after a long-enough time.

From Figs.~\ref{bn}(d), ~\ref{bn}(e), and ~\ref{bn}(f), it is seen that the average particle numbers increase exponentially with time but without oscillations in the region (1). Although the average particle numbers from spontaneous generation play an important role in the total number of particles, it does not dominate the total generation of the average particle numbers. We can see from Eq.~(\ref{spbnb}) that the contribution of spontaneous emission decreases as the initial value increases in this case. Our findings indicate that $\mathcal{PT}$-symmetric optomechanical devices can serve as a powerful tool for controlling photons and phonons.

\begin{figure*}[tbp]
\centering
\includegraphics[width=1 \textwidth]{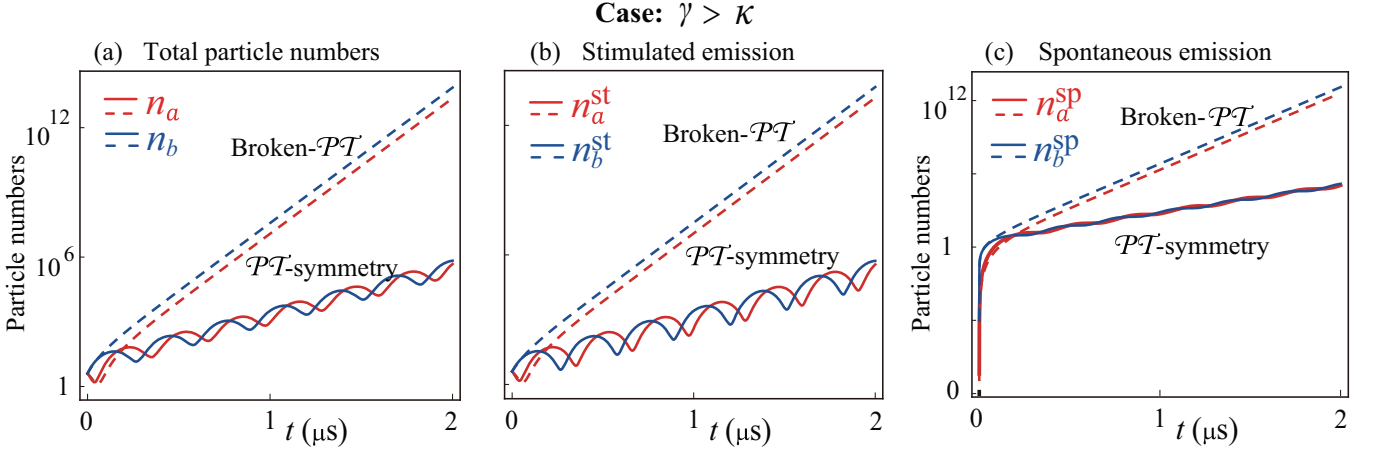}
\caption{Dynamics of photons (red curves) and phonons (blue curves) for the $\mathcal{PT}$-symmetric (solid curves) and broken-$\mathcal{PT}$-symmetric (dashed curves) regimes in the case of $\gamma>\kappa$. (a) The total average particle numbers $n_{a}$ and $n_{b}$, (b) the particle numbers generated by stimulated emission, $n_{a}^{\mathrm{st}}$ and $n_{b}^{\mathrm{st}}$, and (c) those generated by spontaneous emission, $n_{a}^{sp}$ and $n_{b}^{sp}$. Here, the $\mathcal{PT}$-symmetric case is shown for $\gamma=1.8\kappa$ and $G=2.1\kappa$, and the broken-$\mathcal{PT}$-symmetric case is shown for $\gamma=1.8\kappa$ and $G=1.2\kappa$. Other parameters are same as in Fig.~\ref{ex}.}
\label{gn}
\end{figure*}

\subsection{Dynamics of the average particle numbers for $\gamma\neq\kappa$}

Now, we investigate the dynamical behavior of the average particle numbers, $n_{a}$ and $n_{b}$, when $\gamma\neq\kappa$. The expressions of the average particle numbers are given by
\begin{eqnarray}
n_{a}^{\mathrm{st}}&=&\frac{E_t}{d\Omega^{2}}\left(m_{2}+2l_{1}C+2l_{2}S\right),\notag\\
n_{b}^{\mathrm{st}}&=&\frac{E_t}{d\Omega^{2}}\left(m_{2}+2l_{3}C+2l_{4}S\right),\notag\\
n_{a}^{\mathrm{sp}}&=&\frac{4\gamma G^{2}}{d\Omega^{2}}E_t\left[(\gamma-\kappa)^{2}C
-\Omega(\gamma-\kappa)S-4f\right]-\frac{4\gamma G^{2}}{d},\notag\\
n_{b}^{\mathrm{sp}}&=&\frac{4\gamma(\gamma-\kappa)G^{2}}{d\Omega^{2}}E_t\Bigg[\left(\gamma-\kappa-\frac{\kappa\Omega^{2}}{G^{2}}\right)C-\frac{4f}{(\gamma-\kappa)}\notag
\\
&&-\frac{\kappa^{2}-f}{G^{2}}\Omega S\Bigg]-\frac{4}{d}\gamma(\kappa^{2}+f),\label{spnb}
\end{eqnarray}
where $E_t=\exp[(\gamma-\kappa)t]$, $C=\cosh\left(\Omega t\right)$, and $S=\sinh\left(\Omega t\right)$. Similarly to the former case, $\Omega$ is imaginary in the $\mathcal{PT}$-symmetric regime, and the terms $C$ and $S$ transform into the form of sinusoidal time function; while $\Omega$ is real in the broken-$\mathcal{PT}$-symmetric regime and the expression remains the same. Other coefficients are:
\begin{eqnarray}
d&=&4(\gamma-\kappa)f,\notag\\
m_{2}&=&4f\left[i(\gamma^{2}-\kappa^{2})\delta+2G^2(\kappa-\gamma)(|\alpha|^2+|\beta|^2)\right],\notag\\
l_{1}&=&2(\gamma-\kappa)f\left[(\Omega^{2}+2G^{2})|\alpha|^2+2G^{2}|\beta|^2-i(\kappa+\gamma)\delta\right],\notag\\
l_{2}&=&2\Omega(\gamma-\kappa)f\left[i\delta-(\kappa+\gamma)|\alpha|^2\right],\notag\\
l_{3}&=&2(\gamma-\kappa)f\left[2G^{2}|\alpha|^2+(\Omega^{2}+2G^{2})|\beta|^2-i(\kappa+\gamma)\delta\right],\notag\\
l_{4}&=&2\Omega(\gamma-\kappa)f[(\kappa+\gamma)|\beta|^2-i\delta].
\end{eqnarray}
When $\gamma<\kappa$ and $f>0$, the system lies in the asymptotically stable regime. Meantime, the parameters satisfy the relation $G^{2}>(\gamma+\kappa)/2$, the system is $\mathcal{PT}$-symmetric corresponding to the regime (4), which is illustrated by Figs.~\ref{ln}(a),~\ref{ln}(d), and~\ref{ln}(g). These figures shows that the total average particle numbers oscillate in different phase regimes in a certain interval after which it asymptotically approaches an equilibrium value. The oscillation behavior is mainly contributed by the average particle numbers of stimulated generation, while the equilibrium values are only determined by spontaneous generation.

On the other hand, if the parameters satisfy the relation $G<(\gamma+\kappa)/2$, the system is in the broken-$\mathcal{PT}$-symmetric regime (3), which is illustrated by Figs.~\ref{ln}(b),~\ref{ln}(e), and~\ref{ln}(h). Here, the average particle numbers of stimulated generation starts to increase with time, and then decreases to zero; while the average particle numbers due to spontaneous generation increase with time and reach an equilibrium value.
When $\gamma<\kappa$, $f<0$, and $G<(\gamma+\kappa)/2$, the system is in the unstable and broken-$\mathcal{PT}$-symmetric regimes [the regime (1)], which are shown in Figs.~\ref{ln}(c), ~\ref{ln}(f), and ~\ref{ln}(i). The average particle numbers increase exponentially with time, and the spontaneous generation plays an important role only in the total average particle numbers. From Eq.~(\ref{spnb}) we find that the contribution from spontaneous emission decreases with the initial values.

When the parameters satisfy the relation $\gamma>\kappa$, the system is always unstable, the average particle numbers $n_{a}$ and $n_{b}$ have periodic oscillation and their amplitudes increase exponentially with time in the $\mathcal{PT}$-symmetric regime (2), while the oscillation disappears in the broken-$\mathcal{PT}$-symmetric regime (1), which are shown in Fig.~\ref{gn}, respectively. The effect of spontaneous generation on the average particle numbers also decreases with the initial values when $\gamma>\kappa$.

\begin{figure}[tbp]
\center
\includegraphics[width=0.47 \textwidth]{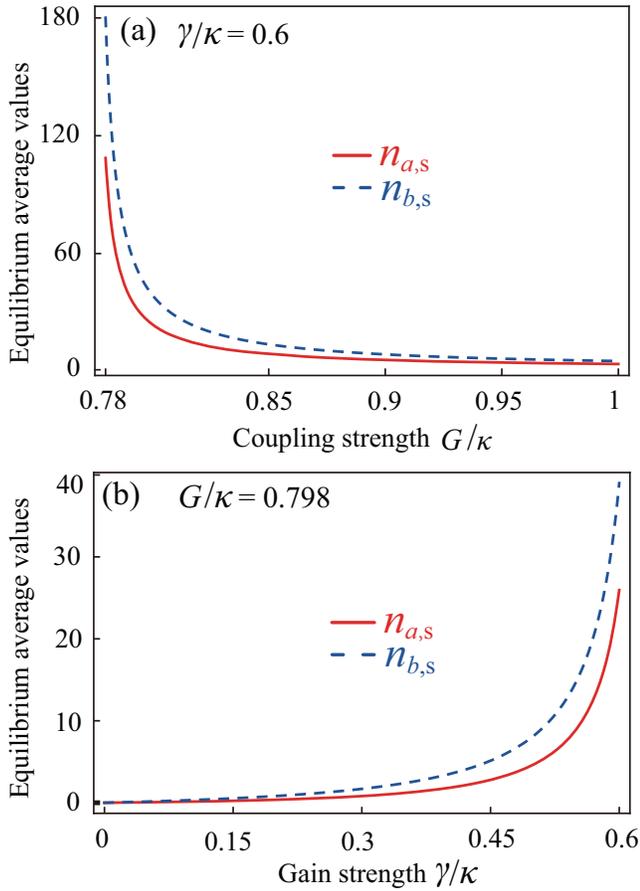}
\caption{(a) Equilibrium average values $n_{a,\mathrm{s}}$ (solid red curves) and $n_{b,\mathrm{s}}$ (dashed blue curves) with respect to the dimensionless normalized coupling strength $G/\kappa$ when $\gamma=0.6\kappa$, (b) Equilibrium values of $n_{a,\mathrm{s}}$ (solid red curve) and $n_{b,\mathrm{s}}$ (dashed blue curve) with respect to the dimensionless normalized gain strength $\gamma/\kappa$ when $G=0.798\kappa$.}
\label{ev}
\end{figure}

We also consider the steady behavior of the average particle numbers $n_{a}$ and $n_{b}$, which are given by
\begin{subequations}
\begin{align}
n_{a,\mathrm{s}}&=\frac{G^{2}\gamma}{(\kappa-\gamma)f},\label{nas}\\
n_{b,\mathrm{s}}&=n_{a,\mathrm{s}}+\frac{\kappa\gamma}{f}.\label{nbs}
\end{align}
\end{subequations}
From Eqs.~(\ref{nas}) and (\ref{nbs}), the equilibrium values $n_{a,\mathrm{s}}$ and $n_{b,\mathrm{s}}$ are independent of their initial values. We consider the variations of the steady values with the normalized coupling strength $G/\kappa$ and normalized gain strength $\gamma/\kappa$, which are shown in Figs.~\ref{ev}(a) and \ref{ev}(b). It is seen that the steady-state values $n_{a,\mathrm{s}}$ and $n_{b,\mathrm{s}}$ decrease with $G$ and approach $\gamma/(\kappa-\gamma)$, while the $n_{a,\mathrm{s}}$ and $n_{b,\mathrm{s}}$ increase with $\gamma/\kappa$.

\section{Conclusions}\label{section:conclusion}

In summary, we have theoretically investigated the dynamics of the average numbers of particles (i.e., photons and phonons) and the average value of the displacement of the mechanical resonator for a $\mathcal{PT}$-symmetric-like optomechanical system. The analytical expressions of these quantities were obtained from the master equation in the full quantum regime, including quantum noise. The dynamics of the number of particles and displacement in different regimes have shown the following characteristics of each regime:

(i) In the $\mathcal{PT}$-symmetric regime, the energy is exchanged rapidly between the cavity and the mechanical oscillator. Moreover, the periodic collapse and revival of the average displacement and the oscillations of the average particle numbers were obtained. In contrast to this regime, all of the studied averages disappear in the broken-$\mathcal{PT}$-symmetric regime.

(ii) In the asymptotically stable regime, the average displacement and the average particle numbers reach their equilibrium values after some evolution time. The average displacement oscillates periodically around zero, and the average particle numbers also oscillate with a monotonously increasing equilibrium value in the finite-time stable regimes (5) and (6), but not asymptotically stable. In the unstable regime, both average particle numbers and displacement increase exponentially.

(iii) Spontaneous emission does not only play an important role for the case of $\gamma=\kappa$, but also for the case of $\gamma\neq\kappa$. And this emission dominates the total generation of the average particle numbers after a long enough time in the finite-time stable regime regime even in the asymptotic limit, while not in the unstable regime. Otherwise, the contribution of spontaneous emission decreases with the initial values.

These results indicate that $\mathcal{PT}$-assisted optomechanical devices can provide a versatile platform to manipulate the mechanical motion, photons, and phonons.

\begin{acknowledgments}
We thank Prof. Hui Jing for his very useful comments. B. P. H. is supported in part by National Natural Science Foundation of China (Grant No.~11974009). A.M. is supported by the Polish National Science Centre (NCN) under the Maestro Grant No. DEC-2019/34/A/ST2/00081. F.N. is supported in part by:
 Nippon Telegraph and Telephone Corporation (NTT) Research,
 the Japan Science and Technology Agency (JST) [via
 the Quantum Leap Flagship Program (Q-LEAP) program,
 the Moonshot R\&D Grant Number JPMJMS2061, and
 the Centers of Research Excellence in Science and Technology (CREST) Grant No. JPMJCR1676],
 the Japan Society for the Promotion of Science (JSPS)
 [via the Grants-in-Aid for Scientific Research (KAKENHI) Grant No. JP20H00134 and the
 JSPS¨CRFBR Grant No. JPJSBP120194828],
 the Army Research Office (ARO) (Grant No. W911NF-18-1-0358),
 the Asian Office of Aerospace Research and Development (AOARD) (via Grant No. FA2386-20-1-4069), and
 the Foundational Questions Institute Fund (FQXi) via Grant No. FQXi-IAF19-06.
\end{acknowledgments}

\end{document}